\documentclass[12pt]{article}
\usepackage{graphicx}

\newcommand\pubnumber{SNSN-323-63}
\newcommand\pubdate{\today}

\def\Title#1{\begin{center} {\Large #1 } \end{center}}
\def\Author#1{\begin{center}{ \sc #1} \end{center}}
\def\Address#1{\begin{center}{ \it #1} \end{center}}

\newcommand\pubblock{\rightline{\begin{tabular}{l} \pubnumber\\
         \pubdate  \end{tabular}}}
\newenvironment{Abstract}{\begin{quotation}  }{\end{quotation}}
\newenvironment{Presented}{\begin{quotation} \begin{center} 
             PRESENTED AT\end{center}\bigskip 
      \begin{center}\begin{large}}{\end{large}\end{center} \end{quotation}}

\def\beq{\begin{equation}}
\def\eeq#1{\label{#1}\end{equation}}
\def\eeqn{\end{equation}}

\def\beqa{\begin{eqnarray}}
\def\eeqa#1{\label{#1}\end{eqnarray}}
\def\eeqan{\end{eqnarray}}

\let\bar=\overbar

\def\Dslash{\not{\hbox{\kern-4pt $D$}}}
\def\dslash{\not{\hbox{\kern-2pt $\del$}}}

\def\msb{{\bar{\ssstyle M \kern -1pt S}}}

\begin{document}
\begin{titlepage}
\pubblock

\vfill
\Title{Plans for future $B$ factories}
\vfill
\Author{P.~Kri\v zan}
\Address{Faculty of Mathematics and Physics, University of Ljubljana,  Ljubljana, Slovenia}
\Address{J.~Stefan Institute, Ljubljana, Slovenia}
\vfill
\begin{Abstract}
The paper  discusses future experiments at super $B$ factories. It presents the physics motivation and the tools, accelerators and detectors, and reviews the status of the two projects, SuperKEKB/Belle-II in Japan and SuperB in Italy.
\end{Abstract}
\vfill
\begin{Presented}
CKM2010, the 6th International Workshop on the CKM Unitarity Triangle, University of Warwick, UK, 6-10 September 2010
\end{Presented}
\vfill
\end{titlepage}
\def\thefootnote{\fnsymbol{footnote}}
\setcounter{footnote}{0}

\section{Introduction}

The two $B$ factories, PEP-II with BaBar and KEKB with Belle, have been a real success story. They 
were built with the primary goal of measuring 
CP violation in the $B$ system. From the discovery of large CP violation in 2001, the $B$ factory results evolved into a 
precision measurement of the CP violation parameter 
$\sin{2\phi_1} = \sin{2\beta} =0.655 \pm 0.024$ in $B \to J/\psi K^0$ decays~\cite{belle-sin2phi1,babar-sin2phi1,hfag-sin2phi1}.
The constraints from measurements of angles and sides of the unitarity triangle show a remarkable agreement~\cite{ckmfitter,utfit,stocchi-ckm2010}, 
which significantly contributed to the 2008 Nobel prize awarded
 to Kobayashi and Maskawa. The two  $B$ factories also
observed direct CP violation in $B$ decays, measured rare decay modes of $B$ mesons, 
and observed mixing of $D^0$ mesons. They measured  CP violation in $b \to s$ transitions, thus 
probing new 
sources of CP violation. The study of forward-backward asymmetry in $b \to s l^+l^-$ has by now 
become a powerful 
tool in the search for physics beyond the  Standard Model (SM). 
Both collaborations also searched for lepton flavor violating $\tau$ decays, and, 
last but not least, observed a long list of new hadrons, some 
of which do not seem to fit into the  standard meson and baryon schemes. 
All this was only possible because of the fantastic performance of the accelerators,
much beyond their design values. In the KEKB case, the  peak luminosity reached a world record 
value of $2.1 \times 10^{34} {\rm cm^{-2}s^{-1}}$, exceeding the design value by a factor of more than two. The two collaborations have accumulated data samples corresponding to integrated luminosities of 0.557~ab$^{-1}$ (BaBar) and 1.041~ab$^{-1}$ (Belle). 

While $B$ factories were built to check whether the SM with the CKM matrix is correct, the
next generation of $B$ factories (super $B$ factories) will have to show in which way the SM is wrong.
To search for departures from the Standard model, a $\approx 50$ times bigger data sample of decays
of $B$ and $D$ mesons and $\tau$ leptons is needed, corresponding to an integrated luminosity of 50-75~ab$^{-1}$. 
A substantial upgrade is therefore required both of the accelerator complex as well as 
of the detector~\cite{loi}.
Note, however, that it will be a different world in four years, when the first super $B$ factory
starts to operate; there will be serious competition from the LHCb and BESIII experiments.
Still, $e^+e^-$ colliders operating at (or near) the $\Upsilon(4S)$ resonance 
will have considerable advantages in several classes of measurements, e.g., with final states involving neutral particles ($\gamma, \pi^0$) and neutrinos, 
and will be complementary in many more. 

In what follows we shall first discuss the physics motivation, the accelerators and detectors, 
and then we shall review the status of the two projects, SuperKEKB/Belle-II in Japan and SuperB in Italy.

\section{Physics motivation}

\begin{figure}[htb]
\begin{center}
\includegraphics[width=0.4\textwidth]{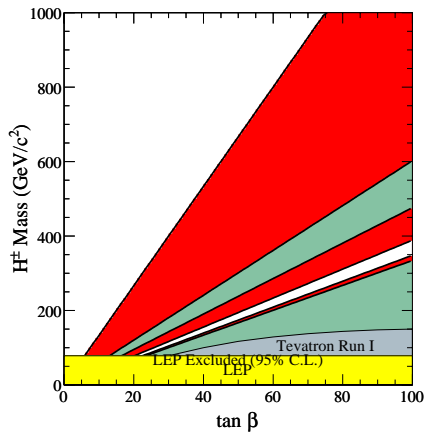}     
\end{center}
\caption{Five standard deviations discovery region (red) for the charged
  Higgs boson in the $(m_{H^\pm},\tan\beta)$ plane, from the
  measurement of ${\cal{B}}(B^+\to \tau^+\nu)$ with
  50~ab$^{-1}$ \cite{phys-belle2}. 
Other shaded regions show the current 95\% C.L. exclusion region.}
\label{fig07}
\end{figure}

Examples of particularly challenging measurements which are only possible at a $B$ factory
 are the studies of $B$ meson decays with more than one neutrino in the final state. 
Such a process is the leptonic decay $B \to \tau \nu_{\tau}$ which is followed by the decay of the $\tau$
lepton with one or two additional neutrinos in the final state. In the SM, this transition proceeds via $W$ annihilation,
but in some new physics (NP) extensions it could also be mediated by a charged Higgs boson~\cite{theo-taunu}. 
 The measured branching fraction can therefore be used to set limits on the  two parameters, 
the charged Higgs mass and the ratio of vacuum expectation values, $\tan{\beta}$.
 As shown in Fig.~\ref{fig07}, with the present measurements (green) it is possible  to exclude a sizable part of the 
parameter space; with a data sample corresponding to a luminosity of 50~ab$^{-1}$, the five standard deviations 
discovery region covers a  substantial fraction of the parameter space (red). The sensitivity is comparable to direct
searches with large data sets at the LHC.

Such rare processes are searched for in the following way~\cite{exp-taunu}. First, one of the $B$ mesons is fully 
reconstructed in a number of exclusive decay channels like $B \to D^{(*)}\pi$. Because of the exclusive
associated production of $B$ meson pairs in a $B$ factory, 
the remaining particles in the 
event must be the decay products of the associated $B$. In the $B^- \to \tau^- \bar{\nu}_{\tau}$, 
$\tau^- \to \mu^- \nu_{\tau} \bar \nu_{\mu}, e^- \nu_{\tau} \bar \nu_{e}, \pi^- \nu_{\tau}, $ decay sequences, 
only one charged particle is detected. To exclude background events with additional neutral particles 
($\pi^0$ or $\gamma$) in the final state,  we use the remaining energy in the calorimeter which is not associated 
with reconstructed charged tracks. In this measurement we greatly profit from the excellent 
hermeticity of the spectrometers of the $B$ factories. 

A similar process, $B \to D \tau\bar{\nu}_{\tau}$ is sensitive to the charged Higgs boson 
as well~\cite{theo-dtaunu}.
Compared to $B \to \tau \nu_{\tau}$, it has a  smaller theoretical uncertainty, 
a larger branching fraction~\cite{exp-dtaunu,dtaunu-review}, and the 
differential distributions can be used to discriminate the contributions of $W^+$ and $H^+$.
It is worth noting that while LHC experiments are sensitive to $H-b-t$ coupling,
in $B \to \tau \nu_{\tau}$ and $B \to D \tau\bar{\nu}_{\tau}$ we probe the $H-b-u$ and $H-b-c$ couplings.

The decay $B \to K^{(*)}   \nu \bar \nu$ has a similar event topology as $B^- \to \tau^- \bar{\nu}_{\tau}$, and a similar event analysis 
can be applied to it as well. By simultaneously measuring the branching fractions for the two decay 
types and comparing them to the SM predictions ($4 \times 10^{-6}$ for $K   \nu \bar \nu$ and 
$6.8 \times 10^{-6}$ for $K^{*}   \nu \bar \nu$, with contribution from penguin box diagrams) 
it is possible to determine the contributions
of anomalous right-handed and left-handed couplings~\cite{altmannshofer,phys-belle2,phys-superb}.

Yet another example of a decay which cannot be studied at LHCb is a measurement of  
CP violation in $B \to K_S \pi^0 \gamma$ decays in a search for right-handed currents.
The present uncertainty in the time-dependent CP violation parameter $S$ is about 0.2, and should be reduced to
 a few percent level with 50~ab$^{-1}$ of data.

Super $B$ factories will also be used to search for lepton flavour violating decays of $\tau$ leptons,
in particular in the $\mu \gamma$ and $\ell \ell \ell$ final state. 
Theoretical predictions for branching fractions of these two decay modes are between $10^{-10}$ and $10^{-7}$ 
for various extensions of SM (mSUGRA+seesaw, SUSY+SO(10), SM+seesaw, non-universal $Z^{0}$, SUSY+Higgs).
The reach of 
super $B$ factories (from $10^{-9}$ to $10^{-8}$, depending on the decay mode) will allow probing of
  these predictions and discrimination between the different NP theories~\cite{goto-07}.

 Two recent publications summarize the physics potential of a super $B$ factory, one prepared by Belle-II authors 
and 
guests~\cite{phys-belle2}, and the other by SuperB collaborators and guests~\cite{phys-superb}. To summarize, 
there is  a good chance to see new phenomena, such as 
CP violation in $B$ decays from new physics sources, or lepton flavor violation in $\tau$ decays.
The Super B factory results will help to diagnose  or constrain  new physics models.  
$B \to \tau\bar{\nu}_{\tau}$ and $B \to D^{(*)}\tau\bar{\nu}_{\tau}$ decays can probe the charged Higgs 
contribution in the large $\tan{\beta}$ region. The physics motivation for a super $B$  factory is independent of LHC. 
If LHC experiments find new physics phenomena, precision flavour physics is compulsory to understand it; 
if no new physics is found at LHC, high statistics $B$ and $\tau$ decays would be a unique way to search for 
new  physics above the TeV scale (or at the TeV scale in case of the minimal flavour violation scenario).
Needless to say that there are many more topics to explore, including CP violation searches for charmed 
hadrons, searches for new hadrons etc.

It is worthwhile to refer to a lesson from history: the top quark mass
was first estimated through the observation of $B^0-\bar{B}^0$ mixing at
ARGUS, and it took seven more years to directly observe it and measure
its mass at the CDF and D0 experiments. Similarly, the prediction of the
charm quark came from the observed absence of flavour changing neutral
currents via the GIM mechanism. Its mass could be estimated from the
observed $K^0$ mixing rate.

\section{Accelerators}

To search for departures from the SM, a $\approx 50$ times larger data sample is needed. 
For such an increase in the data sample, a sizable upgrade of the $B$ factory accelerator complex is required 
 leading to a 40 times larger peak luminosity. These next generation accelerators are known as super  $B$ factories.
There are two super $B$ factory projects under way. The first one, SuperKEKB, foresees a substantial redesign of elements 
of the existing KEKB accelerator complex while retaining  the same tunnel and related infrastructure. 
After 11 years of successful operation, the last  KEKB beam was ceremonially 
aborted on June 30, 2010. This opened the way for the construction of SuperKEKB. To increase the luminosity by a 
factor of 40 the plan is to modestly increase the current (by a factor of 2) with respect to the KEKB values, and dramatically 
shrink the beam size at the collision point, while the beam beam  parameter is kept at the KEKB value (Table~\ref{table1}).
 In this 'nano-beam' scheme which was invented by Pantaleo Raimondi for the Italian SuperB 
project~\cite{nano-beam}, the beams  collide at a rather large angle of 83 mrad (compared to 22 mrad in KEKB). 
In addition, a lower beam asymmetry of 7 GeV and 4 GeV instead of 8 GeV 
and 3.5 GeV is needed to reduce the beam losses due to Touschek scattering in the lower energy beam. 
\begin{table}[t]
\begin{center}
  \begin{tabular}{lccccc}
        \hline \hline
         &                    \multicolumn{2}{c}{SuperKEKB} &       \multicolumn{2}{c}{SuperB} & \\
                              & LER ($e^+$) & HER ($e^-$) & LER ($e^-$)  & HER ($e^+$)   &  \\ 
        \hline
         Energy                & 4.0    & 7.0  & 4.18    & 6.7      & GeV  \\
         Half crossing angle   & \multicolumn{2}{c}{41.5} &   \multicolumn{2}{c}{33} & mrad \\
         Horizontal emittance  & 3.2   & 4.3 & 1.82   & 1.97     & nm \\
         Emittance ratio       & 0.27  & 0.25 & 0.34 & 0.25 & \% \\
         Beta functions at IP (x/y) & 32 / 0.27 & 25 / 0.31  & 32 / 0.205 & 26 / 0.253 & mm \\
         Beam currents         & 3.6 & 2.6 & 1.82 & 1.97 & A \\
         Beam-beam parameter   & 0.0886 & 0.0830  & 0.0970 & 0.0971 & \\
         Luminosity            & \multicolumn{2}{c}{ $8 \times 10^{35}$ } & \multicolumn{2}{c}{ $1 \times 10^{36}$ } & cm$^{-2}$s$^{-1}$ \\

        \hline
\end{tabular}
\caption{SuperKEKB and SuperB: parameters of the low energy (LER) and high energy (HER) accelerator rings.\label{table1}}
\end{center}
\end{table}

The modifications of the  KEKB complex include: improvements in electron injection, 
a new positron target and damping ring,
redesign of the lattices of the low energy (LER) and high energy (HER) rings, 
replacing short  dipoles with longer ones (LER),  
installing TiN-coated beam pipe with ante-chambers, modifications of the RF system,
and a completely redesigned interaction region~\cite{belle2tdr}.

\begin{figure}
\begin{center}
\includegraphics[width=0.7\textwidth]{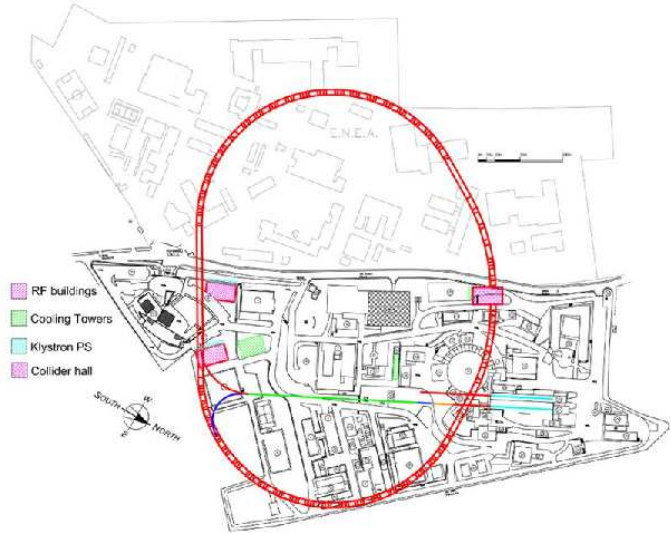} 
\end{center}
\caption{The new SuperB accelerator complex at one of the possible sites, the Frascati National Laboratory.}
\label{superb-lnf}
\end{figure}

Another approach to the design of a super $B$ factory will be exploited in the Italian SuperB project~\cite{superb-acc}.    
Here it is foreseen that a new tunnel will be built (Fig.~\ref{superb-lnf}); the site will be chosen early in 2011. 
Parts of the beam elements of PEP-II will be reused in the accelerator construction. 
In addition to the nano-beam scheme (Table~\ref{table1}), an essential feature of the  SuperB accelerator is 
the crab 
waist collision of two beams in which special sextupoles will be used close to the interaction region 
to maximize the overlap of the two beams. 
This scheme was successfully  tested at the DA$\Phi$NE ring by 
Pantaleo Raimondi and his team~\cite{crab-waist}. 
The SuperB  accelerator is designed in such a way that it can be modified to run at
the $\psi(3770)$ resonance close to charm threshold, where pairs of $D^0$ mesons are produced in a 
coherent $L=1$ state. Data accumulated at charm threshold   would allow precision charm mixing, CP violation 
and  CPT violation studies. Another feature of the SuperB accelerator will be the polarization
of the low energy (electron) beam. This could increase the sensitivity to lepton flavour violating
$\tau$ decays and CP violation in $\tau$ decays through a reduction of backgrounds~\cite{phys-superb}. It would also enable a precise $\sin^2{\Theta_W}$ measurement.

\section{Detectors}

The planned substantial increase in luminosity  requires a careful design of the  detectors. 
To maintain the excellent performance of the spectrometers, the critical issues will be to mitigate the 
effects of higher backgrounds (by a factor of 10 to 20), leading to an increase in occupancy  
and radiation damage, as well as fake hits and pile-up noise in the electromagnetic calorimeter. 
Higher event rates will require substantial modifications in the trigger scheme, DAQ and computing relative to the current experiments.
In addition, improved hadron identification is needed, and  similarly good (or better) hermeticity is 
required~\cite{belle2tdr}.

\begin{figure}
\begin{center}
\includegraphics[width=0.7\textwidth]{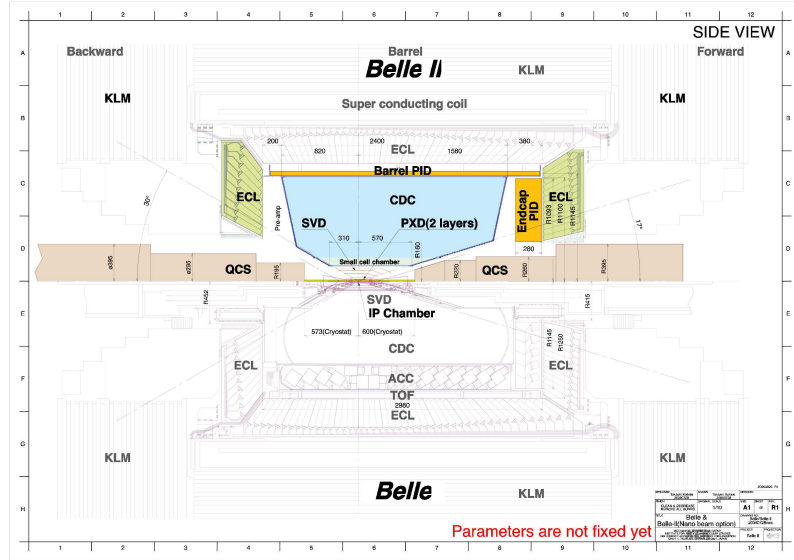} 
\end{center}
\caption{Upgraded Belle II spectrometer (top half) as compared to the present Belle detector (bottom half).}
\label{fig04}
\end{figure}

For the Belle-II detector (Fig.~\ref{fig04}), the following solutions will be adopted~\cite{belle2tdr}. 
The inner layers of the vertex detector
  will be replaced   with a pixel detector, the inner part of the main tracker (CDC, central drift chamber) will be replaced
    with a silicon strip detector, a better particle identification device will be used, the CsI(Tl) crystals
of the end-cap calorimeter will be replaced by pure CsI, the resistive plate chambers of the end-cap muon and 
$K^0_L$ detection system will be replaced by scintillator strips read out by SiPMs, 
and all components will be read-out by 
 fast readout electronics and an improved computing  system. 

The new vertex detector will have two pixel layers, at $r = 14$~mm and $r = 22$~m
 around a 10~mm radius Be beam pipe, and four double-sided strip sensors at  radii of 38~mm, 80~mm, 115~mm, and 140~mm.
The pixel detector will be based on DEPFET sensors~\cite{depfet}.
A significant improvement in vertex resolution is expected with respect to Belle, both for low momentum  
particles (by a factor of two) because of reduced Coulomb scattering, as well as for high momentum particles
because the high resolution pixel detector is closer to the beam pipe and interaction point. Another 
important feature is a significant improvement in $K^0_S$ reconstruction efficiency (by about 30\%)
and vertex resolution because of a larger volume covered by the vertex detector. 

\begin{figure}
\centerline{ 
\includegraphics[width=0.68\textwidth]{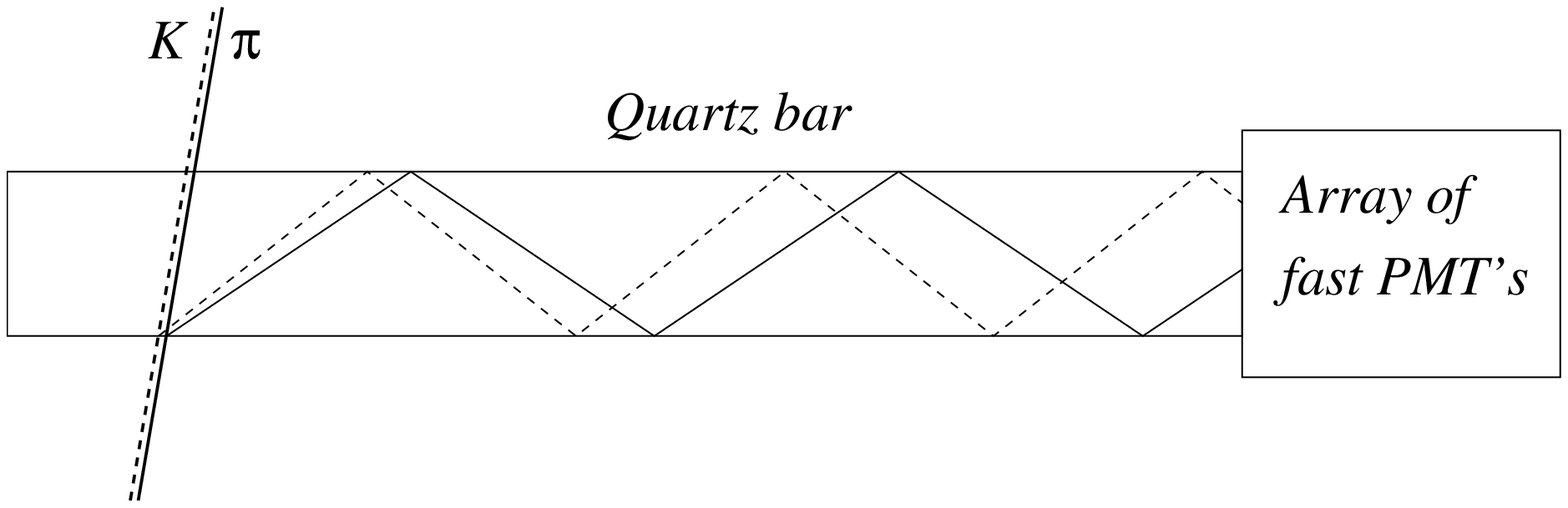}
\includegraphics[width=0.32\textwidth]{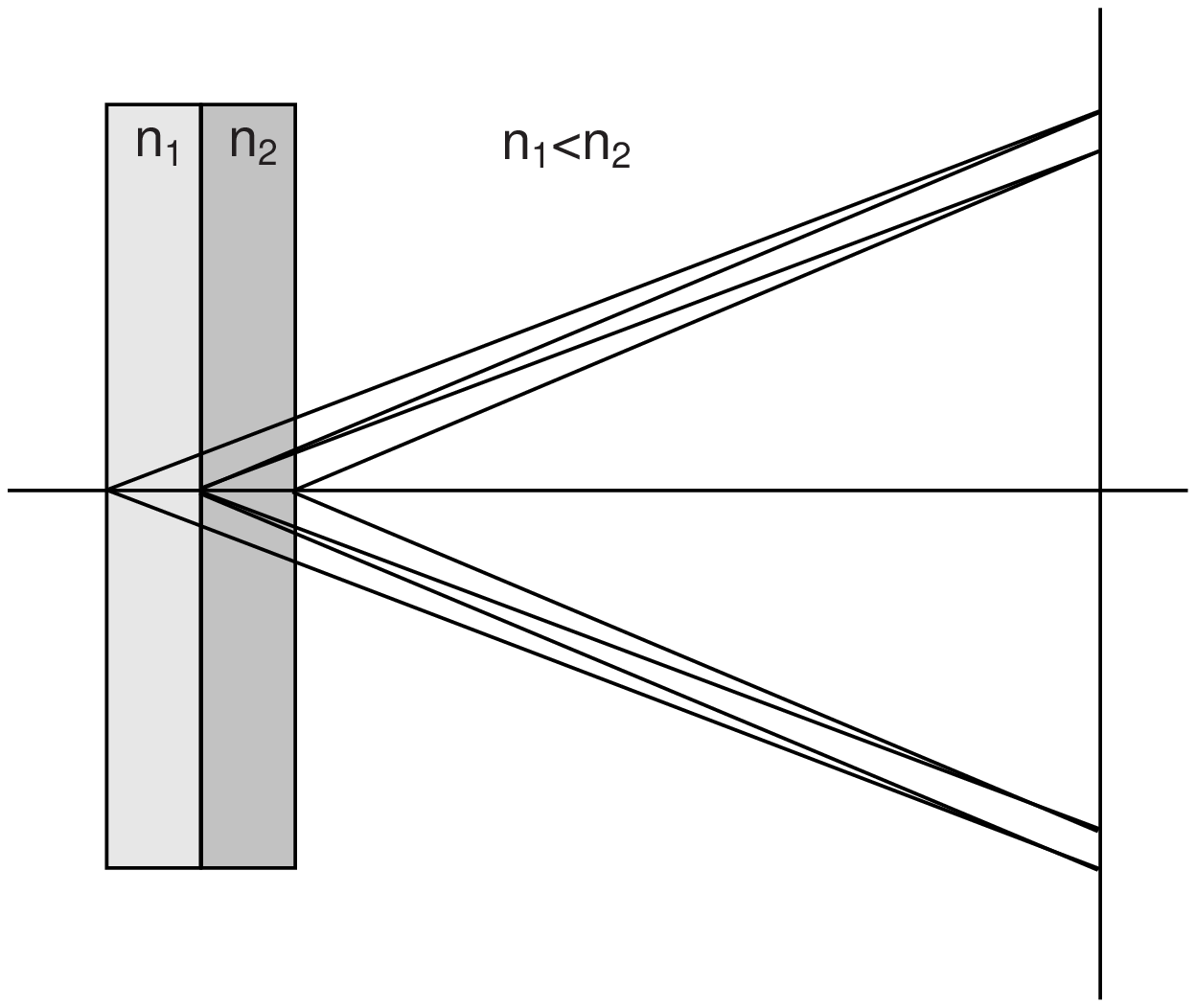}  
}  
\caption{Belle-II PID systems: principle     of operation of the TOP counter (left) and of the 
proximity focusing RICH with  a nonhomogeneous aerogel radiator in the  focusing configuration (right). 
    \label{belle2-pid}}
\end{figure}

 The hadron particle identification will be provided by a time-of-propagation (TOP) counter  
in the barrel part, and a RICH with a focusing aerogel radiator in the forward region of the spectrometer.
The TOP counter~\cite{inami-rich07} is   
a kind of DIRC counter with quartz radiator bars in which the two dimensional information
from a Cherenkov ring image is represented by the time of arrival and 
impact position of the Cherenkov photons at the photon detector. At a given momentum, the slower kaons 
(dotted in Fig.~\ref{belle2-pid}) emit Cherenkov photons at 
a smaller angle than pions; as a result, also their Cherenkov photons propagate longer along the quartz bar. 
Compared to the DIRC, the TOP counter construction is more compact, 
since the large expansion volume is not needed
as the photon detectors can be coupled directly to the quartz bar exit window. 
On the other hand, the TOP counter demands photon detectors with
single photon time resolution below 100~ps.
A 16-channel MCP PMT as developed by Hamamatsu has been investigated
for this purpose~\cite{inami-rich07}.
For the end-cap region a proximity focusing RICH with aerogel as radiator 
is being designed.  
The key issue in the performance of this type of RICH counter is to
improve the Cherenkov angle resolution per track by increasing the number of detected
photons. With a thicker radiator, the number of detected photons increases, but in a
proximity focusing RICH the single photon resolution degrades because of the 
emission point uncertainty. 
However, this limitation can be overcome in a proximity focusing RICH 
with a non-homogeneous radiator~\cite{nim-multi},
 where  one may achieve overlapping
of the corresponding Cherenkov rings on the photon detector (Fig.~\ref{belle2-pid}). 
This represents a sort of focusing
of the photons within the radiator, and eliminates or at least considerably reduces the spread due to emission point uncertainty.  Both detectors are expected to considerably improve the particle identification efficiency if compared to Belle; the end-cap RICH will provide a 4~$\sigma$ $\pi/K$ separation up to kinematic limits, and the barrel TOP counter will identify kaons with an efficiency exceeding 90\% at a few percent pion fake probability.  

The SuperB detector~\cite{superb-det} will reuse several components of the BaBar spectrometer. 
In the baseline version two major changes are foreseen, replacing CsI(Tl) crystals in the forward 
calorimeter with LSO crystals, and a modification of the particle identification device, the DIRC counter.
  Options include a pixel detector layer, a RICH as the forward 
PID device and a veto electromagnetic calorimeter in the backward region to improve the hermeticity of the 
spectrometer. 

\begin{figure}
\centerline{ 
\includegraphics[width=0.9\textwidth]{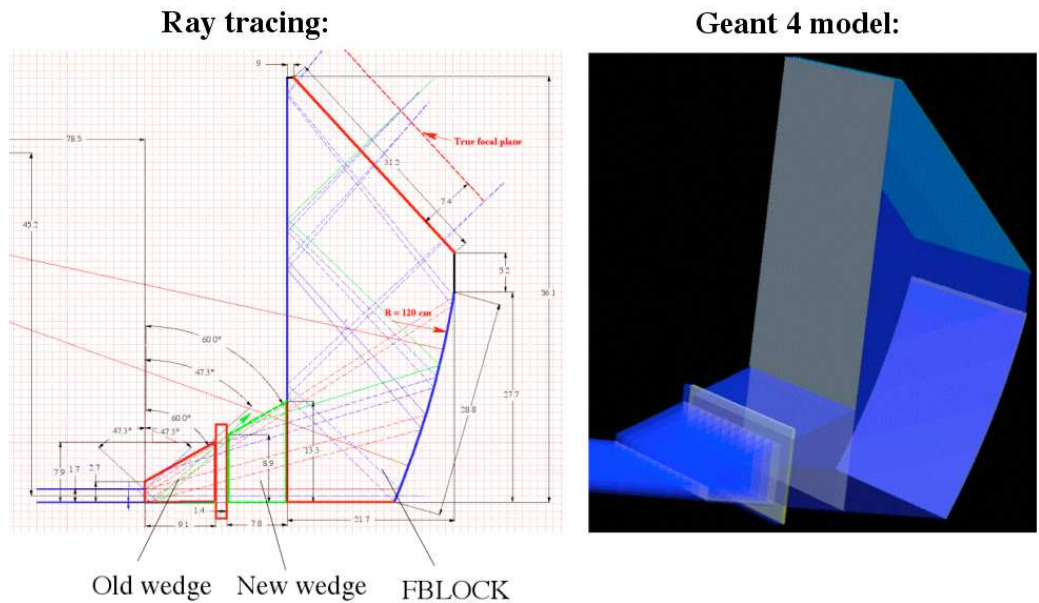}  
}  
\caption{SuperB PID: focusing DIRC counter.
    \label{fdirc}}
\end{figure}
In the new DIRC counter, the large stand-off box with single channel PMTs will be replaced by a compact 
focusing quartz block and multi-anode PMTs as photon sensors (Fig.~\ref{fdirc}). By measuring the time of arrival of Cherenkov
photons, the fast photon detectors will allow to correct for the chromatic error, i.e., variation of 
Cherenkov angle with wavelength~\cite{fdirc}. The focusing DIRC counter is expected to extend the $\pi/K$ separation range by improving the angular resolution by about 10\%. 
At the same time, the order-of-magnitude lower mass of the expansion volume will considerably reduce the level of
beam induced backgrounds.

\section{Status of the projects}

The SuperKEKB/Belle-II project has received initial construction funding in 2010 for the positron 
damping ring, and with the Japanese   'Very Advanced Research Support Program'
a sizable fraction of funds for the main ring upgrade (exceeding 100 MUSD) for 
the period 2010-2012. KEK plans to obtain additional funds to complete the construction as scheduled, i.e., 
start the SuperKEKB commissioning in the autumn of 2014, and start data taking in 2015. It is 
expected that by 2017 the first 5~ab$^{-1}$ of data will be collected, and the full data sample of 
50~ab$^{-1}$ will be reached in 2020/2021. 

The SuperB project  is the first in the list of “flagship projects” of the new Italian national 
research plan over the next few years. The Italian government has delivered an initial funding for 2010 as 
a part of a multi-annual funding program. The aim of the project is to accumulate 75~ab$^{-1}$ 
on a time scale similar to SuperKEKB/Belle-II.

\section{Summary}

$B$ factories have proven to be an excellent tool for flavour physics, with reliable long term operation, 
constant improvement of  operation, achieving and surpassing design performance.
A major upgrade has started at KEK  to construct  the SuperKEKB accelerator and
the Belle-II detector, and be ready for data taking by 2015. The SuperB project  in Italy 
foresees building a new tunnel, reusing and upgrading the PEP-II accelerator and the BaBar detector. Its
special features are a polarized electron beam and the ability to operate at the charm threshold.
Analysis of the physics reach  suggests that we can 
expect a new and exciting era of discoveries, complementary to the LHC.

%
%
 
\end{document}